\documentclass[aps,prl,twocolumn,reprint,amsmath,amssymb,showpacs,superscriptaddress]{revtex4}
\usepackage{graphicx,color,hyperref}
\begin{document}
\def\be{\begin{equation}}
\def\ee{\end{equation}}
\def\bea{\begin{eqnarray}}
\def\eea{\end{eqnarray}}
\def\bef{\begin{figure}[h!]}
\def\eef{\end{figure}}

\def\a{\alpha}
\def\th{\theta}
\def\o{\omega}
\def\eps{\epsilon}
\def\fr{\frac}
\def\l{\label}
\newcommand{\ra}{\rangle}
\newcommand{\la}{\langle}

\title{Comment on: Discontinuous codimension-two bifurcation in a Vlasov equation~\cite{barre2023}.}

\author{Tarc\'isio N. Teles}
\affiliation{Grupo de F\'isica de Feixes, Universidade Federal de Ci\^encias da Sa\'ude de Porto Alegre (UFCSPA), Porto Alegre, RS, Brazil}

\author{Renato Pakter}
\author{Yan Levin}
\email{levin@if.ufrgs.br}
\affiliation{Instituto de F\'isica, Universidade Federal do Rio Grande do Sul (UFRGS), Porto Alegre, RS, Brazil}

\date{\today}

\begin{abstract}
\end{abstract}

\date{\today}
\maketitle


In their recent work~\cite{barre2023}, Yamaguchi and Barr\'e (YB) conducted a stability analysis of a class of initially homogeneous solutions to the Vlasov equation for a generalized Hamiltonian mean field (gHMF) model  -- a system of unit-mass particles confined to a ring and interacting via a \(2\pi\)-periodic pair potential, \(\phi(\theta) = -[K \cos(\theta) + K_2 \cos(2\theta)]\), where \(K\) and \(K_2\) represent the strengths of the two potential terms, and \(\theta = \theta_i - \theta_j\) denotes the angular separation between two arbitrary particles \(i\) and \(j\). The authors investigated the stability of a family of stationary paramagnetic states (in which particles are uniformly distributed between $(0, 2\pi]$, with the one particle distribution function given by $F_{\alpha}(\theta, p) = A \exp\left(-\frac{\beta_2 p^2}{2} - \frac{9p^4}{4}\right)$, where \( A \) is the normalization constant satisfying \(\iint F_{\alpha}(\theta, p) \, d\theta \, dp = 1\) and \(\alpha = -\beta_2 A\). These distributions are classified as unimodal for \( \alpha \leq 0 \) and bimodal for \( \alpha > 0 \). Henceforth referred to as the momentum distribution function, as illustrated in Fig. 2 of~\cite{barre2023} and reproduced in Fig.~\ref{fig0} below.

\begin{figure}[!h] 
\centering 
\includegraphics[scale=1.25, width=7cm]{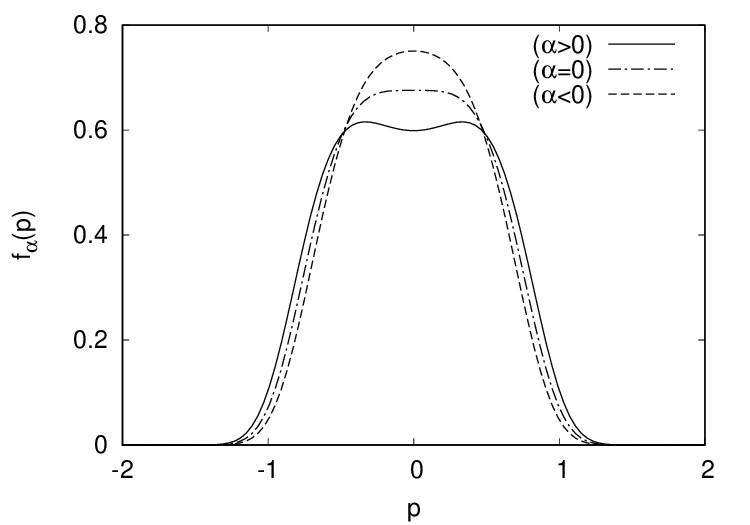} 
\caption{Momentum distribution functions for the unimodal, flat-top, and bimodal cases, where \( f_{\alpha}(p) = \int F_{\alpha}(\theta, p) \, d\theta \). The distribution differ in terms of their concavity at the origin.} 
\label{fig0} 
\end{figure}

Using linear perturbation theory, the authors identified the stability threshold at which the momentum distribution becomes unstable, referring to it as the ``critical point",  thus drawing an analogy with continuous equilibrium phase transitions~\cite{ogawa1}.
They state:  \textit{``...bifurcations have a universal character and tend to provide information about the structure of the phase space, sometimes extending over a rather wide neighborhood of the critical point."}  Indeed, the belief that ``bifurcation" leads to a phase transition is common in the literature on long-range interacting systems, see for example~\cite{bachelard}.


Nevertheless, a claim that the bifurcation analysis alone can provide a meaningful insight into the nature of a quasi-stationary state (qSS) to which the system will evolve,  contradicts our recent findings~\cite{arxiv}.
To explore this issue in greater depth, we conducted extensive molecular dynamics simulations, which demonstrate that the bifurcation analysis is insufficient to predict either the location or the order of the phase transition between the qSS states.

The gHMF model used by YB to justify their assertions was introduced in our earlier work~\cite{prl, pre, report}  and was more recently analyzed in~\cite{marciano}. The gHMF model combines ferromagnetic and nematic-like interactions, providing a versatile framework for exploring complex phase behaviors and non-equilibrium dynamics. Moreover, it serves as a long-range extension of the generalized \(XY\)-model, initially introduced in its short-range form in~\cite{Lee1985}. 
While in our previous study~\cite{prl, report}, the parameter \(K_2\) was set as \(K_2 = 1 - K\), with \(K\) varying within the range \(K \in [0, 1]\), YB, in their work~\cite{barre2023}, fixed \(K_2 = 0.5\) and used an arbitrary value of \(K\) in conjunction with the parameter \(\alpha\) to identify the instability threshold at which the uniform (paramagnetic) particle distribution becomes unstable.  Furthermore, unlike our earlier work on the gHMF model, which used the water-bag distribution for initial particle position and velocity distribution, YB studied more general initial velocity distributions presented in Fig. 1.  


It is well established that for systems with long-range interactions, in the limit of large number of particles, MD simulations become equivalent to the Vlasov evolution on a suitably fine grid~\cite{Braun:1977}. In this study, we simulate a system comprising \(N = 10^8\) particles, each evolving under Hamilton's equations, subject to the same interaction potential \(\phi(\theta)\) considered by YB. To maintain consistency with the notation used in~\cite{barre2023}, we enforce the symmetry condition for the initial distribution, so that \(m_{l,y} = 0\) for \(l = 1, 2\). Furthermore, since \(K > K_2 = 0.5\), the relevant order parameter is \(m_{1,x}\), which we simply denote as \(m(t) = \langle \cos[\theta(t)] \rangle\), where \(\langle \cdot \rangle\) represents the average over all particles.

For the unimodal case with \(\alpha = 0\), the results align with our earlier findings for the flat-top water bag (WB) distribution, predicting a discontinuous transition between the paramagnetic and ferromagnetic states. We next focus on the bimodal ($\alpha>0$) case studied by YB, specifically the one presented in Fig. 7, panel (c) of~\cite{barre2023}. The theory developed by YB predicts that the transition at the instability threshold is continuous and occurs at \(K \approx 0.95\). When analyzing the time series for magnetization, we indeed observe that as the threshold is crossed the amplitude of oscillation of the order parameter $m(t)$ starts to increase, however the time average ${\overline m}$ remains zero, as is shown in Fig.~\ref{fig3} below.
\begin{figure}[!h] 
\centering 
\includegraphics[scale=1.5, width=7.cm]{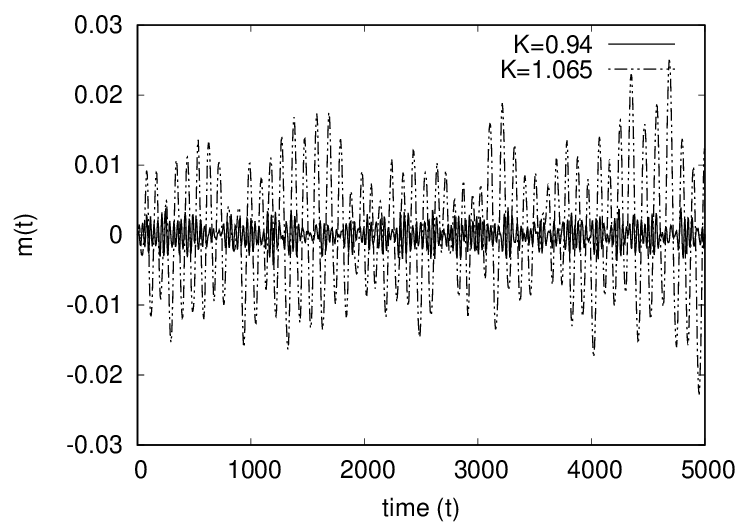} 
\caption{Magnetization before and after the bifurcation point $(K_c=0.95)$  identified by YB for the bimodal distribution with  \(\alpha = 0.0311643\). In both instances, the time-averaged magnetization is zero  -- indicating that the system remains in a paramagnetic state above the instability threshold.} 
\label{fig3} 
\end{figure}

\begin{figure}[!h] 
\centering 
\includegraphics[scale=1., width=7.cm]{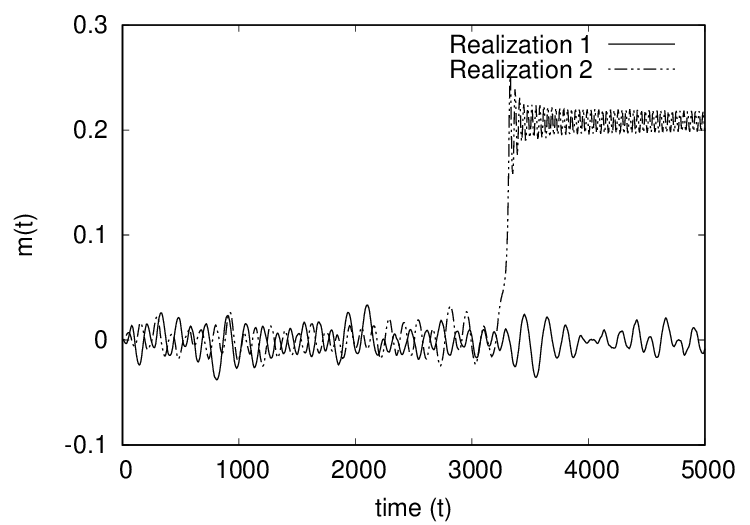} 
\caption{For \(K = 1.070\), corresponding to the bimodal distribution with \(\alpha = 0.0311643\), we observe a coexistence of two distinct qSS using particles with exactly the same initial momentum  and differing only by a random number exchange in the homogeneous angular distribution. In all cases, the initial magnetization was \(|m_0| < 10^{-6}\), and the particles evolve using a fourth-order symplectic integrator~\cite{Yoshida1990} with a time step of \(dt = 0.02\). The energy conservation in the simulation is maintained with a precision accurate to the seventh decimal place.}
\label{fig2}
\end{figure}

Therefore, after the instability threshold, the system remains paramagnetic! The instability of the initial distribution does not necessarily imply a phase transition  -- only that the initial paramagnetic distribution becomes unstable and will oscillate. 
On the other hand YB incorrectly interpret the oscillating state as indicative of a continuous transition to a ferromagnetic qSS and attempt to use the amplitude of the oscillation as the order parameter to extract a scaling exponent, analogous to the critical exponents in equilibrium systems~\cite{ogawa2}. 
In view of our MD simulation results, this interpretation is clearly flawed, since the time-averaged magnetization remains zero both above and below the instability threshold.  The real paramagnetic-ferromagnetic transition is observed to be of first order and occurs at  $K$ significantly 
larger than the threshold $K_c$ predicted by YB.  
Indeed, as we increase the parameter \(K\) further, we begin to observe a coexistence of two distinct qSS: one ferromagnetic and the other oscillating (paramagnetic), as illustrated in Fig.~\ref{fig2}.  
The system evolves to either one of these qSS,  starting from initial conditions drawn from {\it exactly} the same distribution.  This ``coexistence" is a clear hallmark of a first order phase transition. As we increase $K$ even further we leave the coexistence region, and observe that all the initial conditions drawn from the same distribution evolve to a ferromagnetic qSS.
\begin{figure}[!h]
\centering 
\includegraphics[scale=1.,width=7.cm]{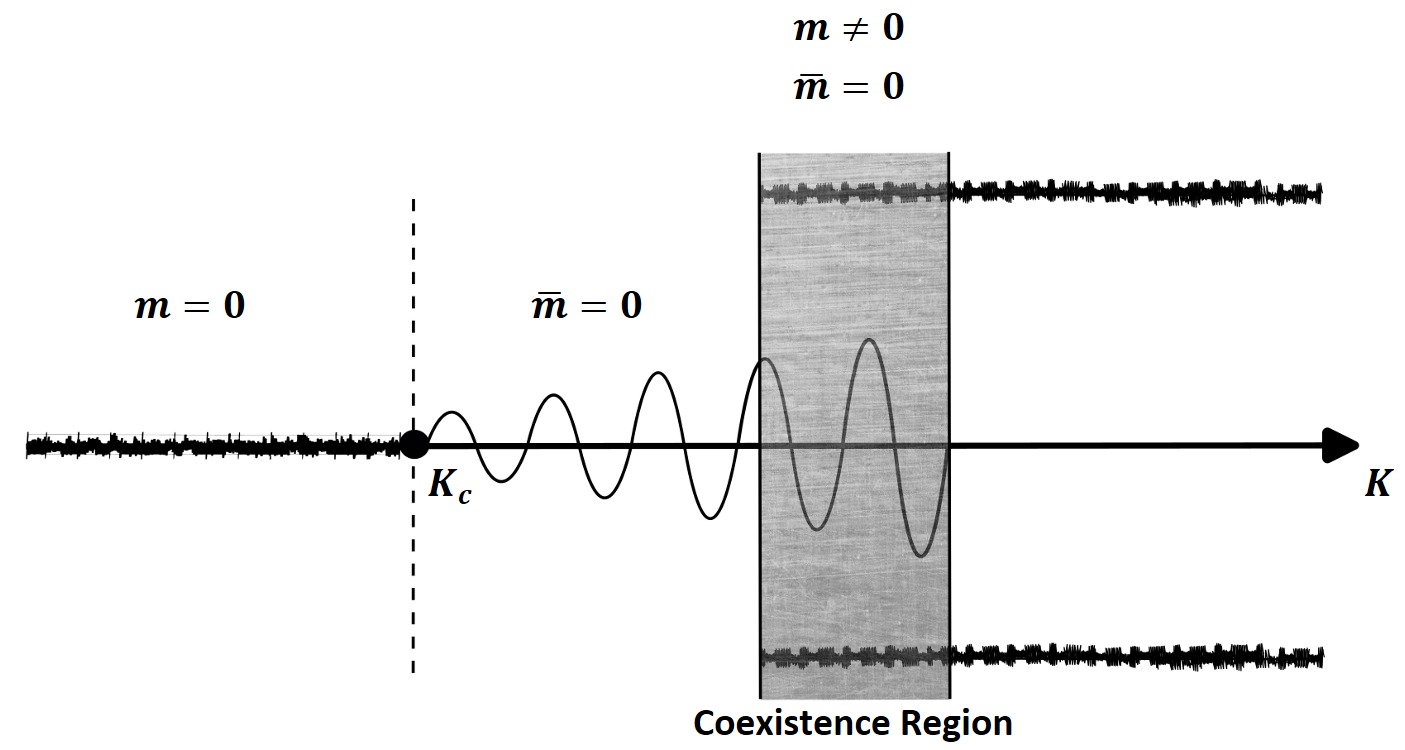} 
\caption{Schematic representation of the system's behavior as a function of the parameter \(K\) for the bimodal case (\(\alpha > 0\)). Below the threshold \(K_c\), the magnetization exhibits microscopic fluctuations around zero. Beyond the threshold, the system develops oscillations around  $\overline{m}=0$. For larger $K$ we reach a coexistence region, in which initial conditions drawn from the same distribution evolve either to a paramagnetic or a ferromagnetic state.   For still larger $K$  all the initial conditions evolve to a ferromagnetic state.  Notably, as \(\alpha \to 0\), the region characterized by the paramagnetic state with oscillating magnetization appears to shrink, causing the lower boundary of the coexistence region to move towards \(K_c\). In either cases, we conclude that the system undergoes a discontinuous order-disorder phase transition.} 
\label{fig5} 
\end{figure}
 In Fig.~\ref{fig5} we summarize the phase diagram as a function of $K$.  
In passing we note that the time window employed in the analysis of Ref.~\cite{barre2023} is insufficient to adequately analyze the qSS to which the system evolves. A longer observation period is required to observe the jump in magnetization, which serves as a critical indicator of a discontinuous phase transition. 


In summary, the simple linear stability analysis adopted by YB is insufficient to make any general prediction about the  nature of the qSS to which the system will evolve.  Indeed, for some initial distributions, after the bifurcation, the system may evolve to ferromagnetic states, for other distributions, however, it will remain paramagnetic. The bifurcation analysis conducted by YB does not provide any meaningful insight into the location or the order of the true paramagnetic-ferromagnetic phase transition in the gHMF or, for that matter, even for the simpler HMF model.      
By contrast, the recently developed adiabatic local mixing (ALM) theory~\cite{arxiv} offers accurate predictions for both the location and the order of symmetry-breaking transitions in these systems, going  beyond the linear stability analysis.



\end{document}